\documentclass[aps,prl,twocolumn,showpacs,floatfix]{revtex4}
\usepackage{epsfig}
\begin{document}
\title{A variational approach to the optimized phonon technique for
electron-phonon problems}
\author{V. Cataudella, G. De Filippis, F. Martone, and C. A. Perroni}
\affiliation{Coherentia-INFM and Dip. di Scienze Fisiche, Universit\`{a}
di Napoli ''Federico II'', Comp. Univ. M.S.Angelo Napoli, Italy}
\date{\today}

\begin{abstract}
An optimized phonon approach for the  numerical diagonalization of interacting
electron-phonon systems is proposed.  The variational method is based on
an expansion in coherent states that leads to a dramatic truncation in the
phonon space.  The reliability of the approach is demonstrated for the extended
Holstein model showing that different types of lattice distortions are present
at intermediate electron-phonon couplings as observed in strongly correlated
systems. The connection with the density matrix renormalization
group is discussed.
\end{abstract}


\maketitle

The interaction of electrons with local lattice deformations plays an important
role in many different materials leading to a number of effects that range
from colossal magnetoresistance in manganites\cite{edwards} to
superconductivity in fullerenes,\cite{fullereni} from pseudo-gap in
cuprates\cite{bishop} to Peierls instability in quasi-one-dimensional
materials,\cite{bray} only to mention a few.
Although the theoretical study of the electron-phonon (e-ph) interaction has
attracted constantly the interest of the scientific community, several
aspects are still not fully understood and challenging. Actually,
analytical solutions of the e-ph problem are available only in
weak and strong coupling asymptotic regimes, even for the simplest molecular
crystal model introduced by Holstein.\cite{holstein}
Some insight in the problem for intermediate
coupling comes from the dynamical field approach\cite{DMF} that provides
exact results in the infinite dimension limit and variational methods.\cite
{noi} In order to achieve a complete understanding of the e-ph interaction,
numerical techniques as exact diagonalization and Quantum Monte Carlo have
been exploited. In both cases the specific problem posed by the
e-ph interaction is related to the presence of phonons that
require an infinite dimensional Hilbert space. As a consequence, for
instance, the direct application of Lanczos exact diagonalization has been
limited to very small lattices and needs a somehow arbitrary truncation in the
phonon number. In this context the introduction of an optimized phonon
approach based on the analysis of the density matrix\cite{white, wellein}
has produced an important improvement. The idea beyond this approach is to
take advantage of the knowledge of the largest eigenvalues and eigenvectors
of the site density matrix to select the phonon linear combinations that, at
the best, can describe the system: the so-called optimized phonon basis
(OPB). Unfortunately, the density matrix of the target states is not
available {\em a priori}: it must be calculated in a self-consistent way
together with the OPB. To this aim different strategies have been discussed.
In this report we wish to introduce a variational technique, based on an
expansion in coherent states, that very much simplifies the selection of an
optimized phonon basis and that does not require any truncation
in the number of phonons.
The method is able to provide accurate results for any e-ph
coupling regime and for any value of the adiabatic ratio. As we will show in
the case of the Holstein model, the proposed expansion provides states
surprisingly close to the eigenvectors of the site density matrix
corresponding to the highest probabilities and allows us to clarify some
questions of the e-ph interaction that have been debated
recently. In particular, the role of quantum lattice fluctuations in the
Holstein model at half-filling and in the clustered phase is discussed
emphasizing the coexistence of very different lattice deformations
that all contribute to the ground state.

{\em The model and the proposed optimized phonon basis. }In the framework of
e-ph systems the Holstein model has been very much studied.
Here we discuss a simple generalization that includes
nearest neighbor e-ph interaction terms. Furthermore, we consider spinless
fermions in order to simulate at a very rough level\cite{nota1} a strong
on-site electron repulsion. In the 1D case the Hamiltonian is

\begin{eqnarray}
H &=& -t\sum_{i}\left( c_{i}^{\dagger }c_{_{i+1}}+h.c.\right) +\omega
\sum_{i}b_{i}^{\dagger }b_{_{i}}\nonumber\\
&+& g\omega\sum_{i}c_{i}^{\dagger
}c_{i} \left[ \left( b_{i}^{\dagger }+b_{i}\right) +\varepsilon
\sum_{\delta =\pm 1}
\left(b_{i+\delta }^{\dagger }+b_{i+\delta }\right) \right]
\label{hamiltonian}
\end{eqnarray}
where $c_{i}^{\dagger }$ ($c_{i}$) is the site electron creation
(annihilation) operator and $b_{i}^{\dagger }$ ($b_{i})$ creates
(annihilates) a phonon in the site $i$. The phonon frequency is $\omega $,
the electron hopping between nearest neighbors is controlled by $t$ while $g$
and $g\varepsilon $ describe the strength of the on-site and nearest neighbor
e-ph interactions, respectively. We assume $\hbar=1$.

The ''natural'' basis in which we can describe Hamiltonian (\ref{hamiltonian})
is given by
\begin{equation}
\left| {\bf \nu },{\bf \mu }\right\rangle =\prod_{i}\left| \nu
_{i}\right\rangle \left| \mu _{i}\right\rangle ,  \label{nb}
\end{equation}
where $i$ run over the lattice sites, $\nu _{i}$ is the electron state label
that, for spinless electrons, can assume only two values and $\mu _{i}$
labels the infinite phonon states on-site $i.$ Our aim is to achieve a very
satisfying estimation of the ground state of (\ref{hamiltonian}) by using
only a finite number of states. The first step is to replace the phonon
states $\left| \mu _{i}\right\rangle $ at site $i$ with coherent states (CS)
\begin{equation}
\left| h,i\right\rangle =e^{gh(b_{i}-b_{i}^{\dagger })}\left| 0\right\rangle
_{i}^{_{(ph)}} ,
\label{CS}
\end{equation}
where $\left| 0\right\rangle _{i}^{_{(ph)}}$ is the phonon vacuum at the
site $i$. This substitution does not introduce any truncation in the Hilbert space
since, varying $h$ in the complex plane, the local basis (\ref{CS}) is
over-complete. The second step is to choose a finite number $M$\ of CS and
the corresponding values $h_{\alpha }$\ ($\alpha =1,...,M$) that we have
supposed not to depend on the site. The most efficient way to proceed is to
fix the number of CS arbitrarily and choose the $h_{\alpha }$ variationally
calculating the ground state energy and minimizing it. In this way we obtain
the ''best'' sub-space where to find the ground state. The method has the
advantage to allow a systematic improvement of the approximation adding more
and more CS and provides a quantitative estimation of the error introduced
by a specific truncation. Of course, the number of CS per site cannot grow as we
wish since the size of the matrix to be diagonalized can become very large.
For instance, if we choose two CS per site, the phonon matrix at fixed
electron configuration will be of size $L=2^{N}$, where $N$ is the number of
the lattice sites. As we will show, two states per site are a good choice
for a very accurate estimation of the ground state.

The proposed optimized phonon basis  can further be improved introducing a
dependence on the electron state at site $i$ since, on physical ground, the
lattice deformation of site $i$ is expected to depend very much on whether
or not the site is occupied by an electron. This can be done acting with the
operator $S$ of the usual Lang-Firsov transformation\cite{lang-firsov} on
each CS belonging to the chosen basis :
\begin{equation}
S=e^{g\left[ \left\langle n_{i}\right\rangle +f(n_{i}-\left\langle
n_{i}\right\rangle )\right] (b_{i}-b_{i}^{\dagger })},  \label{LF}
\end{equation}
where $f$ depends neither on site $i$ nor on the specific CS to which $S$ is
applied.
The value of $f$ can be estimated variationally together with the set $%
\left\{ h_{\alpha }\right\} $. Technically, it is easier first to work out
the action of $S$ transforming $H$ to $H^{\prime }=S^{\dagger }HS$ and,
then, to study the properties of $H^{\prime }$ in the subspace spanned by
both the electron and the optimized phonon basis:
\[
\left| \left\{ h_{\alpha }\right\} \right\rangle =\prod_{i=1}^{N}\left|
h_{\alpha _{i}},i\right\rangle =\prod_{i=1}^{N}\left[ e^{gh_{\alpha
_{i}}(b_{i}-b_{i}^{\dagger })}\left| 0\right\rangle _{i}^{_{(ph)}}\right],
\]
where $h_{\alpha _{i}}\in \left\{ h_{\alpha }\right\} $ and $\alpha
_{i}=1,...,M$.

\begin{figure}
\begin{center}
\epsfig{figure=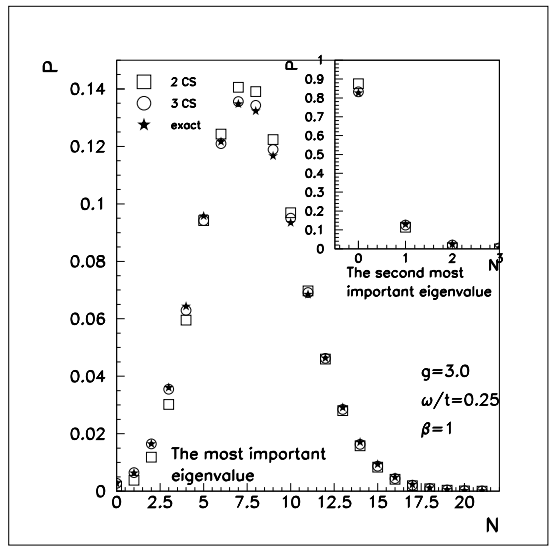,width=0.4\textwidth}
\end{center}
\caption{The probability $P$ of the $N$ phonon state for the  eigenvector
corresponding to the largest eigenvalue of the site density matrix
($\beta =1$, occupied site). The system is made of two lattice sites and
one electron.  The numerically exact results
(stars) are compared to the results obtained within our approach for two
(squares) and three (circles) CS. The values of the parameters
are chosen in the intermediate coupling region (''worst'' case).
In the inset: the second most important eigenvalue.}
\label{Fig.1}
\end{figure}
{\em Single polaron features and site density matrix. }In order to clarify
the relation between the proposed approach and the optimized phonon states
obtained within the density matrix method, we study a two-site one-electron
system.
In particular, we can construct the site density matrix\cite{white}
calculating its eigenvalues and eigenvectors. Both these
quantities can be evaluated both within our approach and by an exact
diagonalization method in the ''natural'' phonon basis.
The comparison between the two
results is contained in Fig.1 where the two eigenvectors corresponding to
the largest eigenvalues are plotted as a function of the phonon number for $%
\beta =1$ (occupied site). It is clear that already with two CS the
agreement is surprisingly good while with three CS the differences with the
numerically exact results are completely negligible. The agreement is due to
the fact that the exact site density matrix contains only a few relevant
eigenvalues and that the corresponding eigenvectors can be approximated as
linear combinations of CS. It is worthwhile noting that, roughly speaking,
in the strong and intermediate coupling regime, the two most important
eigenvectors correspond to strong (weak) lattice deformations\ and weak
(strong) lattice deformations, respectively. The two contributions are both
important in the so-called intermediate regime while in the weak and strong
regimes only one type of lattice deformation is relevant. Also for $\beta
=0$ (empty site) the site deformation is built up through both weak and/or
strong contributions; however, the method proposed is so flexible to allow
different distortions for empty and occupied sites. These observations make
clear why the generalized Lang Firsov method\cite{lang-firsov},
based on a single CS, is not able to give satisfactory results in the
intermediate regime (Fig.1).
One of the advantages of the proposed method is the possibility to
drastically reduce the size of the matrix that represents the system
Hamiltonian (\ref{hamiltonian}). We were able with little computational
effort to calculate the ground state energy of a single
polaron in a 16 site lattice by using two CS. The comparison
with current estimations in the
literature\cite{kornilovitch,feskenew} is excellent (Fig. 2 and Table 1)
showing that two CS are enough to give a correct description in any regime.

\begin{table}
\caption{The ground state energy $E/t$ of a polaron for different regimes
by using $2,3,4$ CS per site is compared with results provided by
Wellein et al.\cite{feskenew}. The system size is $L=10$.}
\begin{tabular}{cccccc}
\hline\hline
$g^2$ & $\omega/t$ & E/t[15]  & 2CS & 3CS & 4CS \\ \hline
$0.5$ & $0.4$ & $-2.06130$ & $-2.06128$ & $-2.06130$ & $-2.06130$ \\
$5.0$ & $0.4$ & $-2.72827$ & $-2.71365$ & $-2.72777$ & $-2.72828$ \\
$1.0$ & $4.0$ & $-5.02985$ & $-5.02883$ & $-5.02984$ & $-5.02985$ \\
\hline\hline
\end{tabular}
\end{table}

\begin{figure}
\begin{center}
\epsfig{figure=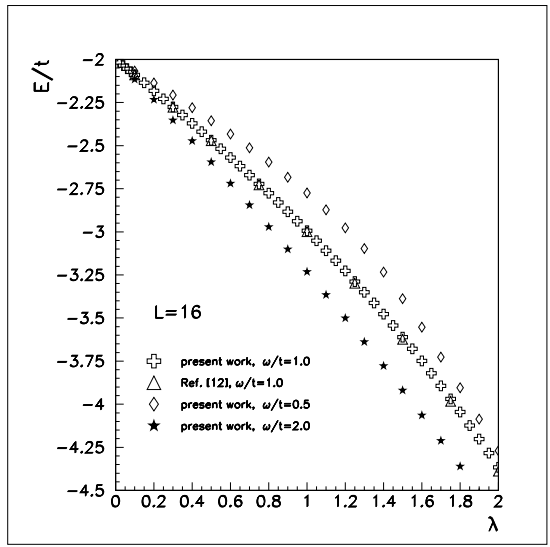,width=0.4\textwidth}
\end{center}
\caption{The ground-state energy of a polaron as a function of the
coupling constant $\lambda=g^2\omega/2t$ for different adiabatic ratios
$\omega/ t$. The case $\omega/ t=1$ is compared with the data of
Ref.\cite{kornilovitch} }
\label{Fig.2}
\end{figure}
{\em The half-filling case: anti-adiabatic regime and quantum fluctuations.}
It is well known that the ground state of the 1D Holstein model exhibits a
phase transition from a Luttinger liquid to a charge ordering state at
half-filling, \cite{CO} but the role of the quantum fluctuations has not
been fully explored. First of all we have checked that the total energy and
the static structure factor calculated with our method are in excellent
agreement with previous estimations based on exact diagonalization obtained
by Lanczos method.\cite{feskeold} In particular, in the intermediate regime
(the worst case), our estimations based on two CS is $0.1\%$ higher than the
best estimation available.\cite{feskenew} In order to understand the role of
the quantum fluctuations in the anti-adiabatic regime ($\omega \gg t$) where
quantum lattice effects are expected to play an important role, we have
calculated $P(X;\chi)$, the lattice displacement probability distribution
function (LDPDF), at a fixed electron configuration $\chi $.
In Fig. 3 we show the results of the calculated LDPDF, relative to a
specific site, associated to the most important electron configurations
close to the boundary of the CO phase for $\omega =5t$ and $g^2=3.6$. The
full line gives the LDPDF of an occupied site in the CO electron
configuration where alternating sites are occupied. As expected it is peaked
at a large value of the lattice displacement (roughly $X_{0}\sqrt{2M\omega }%
\simeq 2g\simeq 3.8$, $M$ being the ionic mass)
since we are in the strong coupling regime.
However, in Fig. 3 it is shown that there are less
important electron configurations among which some contribute to the
LDPDF with a peak
at around $X_{0}\sqrt{2M\omega }\simeq 0$ typical of empty sites. These less
important electron configurations can be viewed as ''defects'' with respect
to the ''ideal'' CO configuration. These defects are associated to very
different lattice displacements compared to the ones characteristic of the
dominant CO configuration, but are still important in the anti-adiabatic
regime. We expect that this behavior is still present in the thermodynamic
limit since the electron configurations shown in Fig.3 are relevant also
in the broken symmetry phase of the thermodynamic limit.
We note that the strong fluctuations
represent the quantum counterpart of the thermal fluctuations found in the
opposite adiabatic limit within the dynamical mean field approach (DMF).\cite
{ciuchi} Furthermore, we stress that, in the adiabatic limit, the quantum
fluctuations behave in a very different way giving rise to ''defects'' whose
lattice displacement is of the same type of the dominant CO configuration.
For this reason, as shown in DMF\ approach, the ground state LDPDF is
characterized by a single peak.

\begin{figure}
\begin{center}
\epsfig{figure=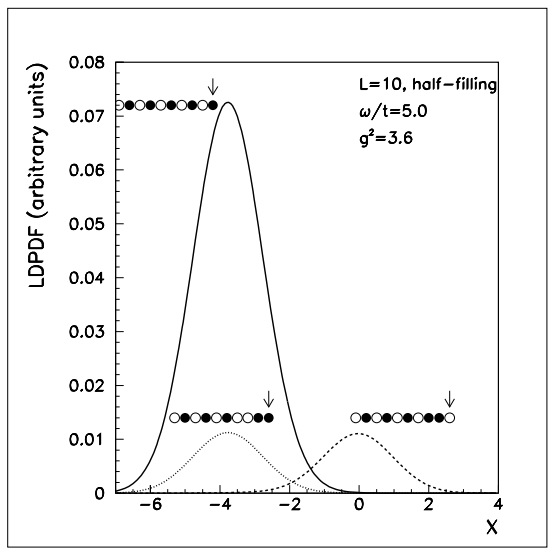,width=0.4\textwidth}
\end{center}
\caption{The LDPDF of the site indicated by the arrow for the most important
electron configurations are
plotted as a function of the lattice distortion in the anti-adiabatic
region and in the charge-ordering regime.
There are important electron configurations characterized by very
different distortions. The lattice distortions are measured in units of
$(2M\omega)^{-1/2}$.}
\label{Fig.3}
\end{figure}

{\em Evidence of quantum fluctuations in clustered phases at intermediate
coupling.} Recently the formation of clusters due to e-ph interaction has
been discussed in connection to experimental evidences of charge ordering in
cuprates and manganites.\cite{tranq,cheong} One of the simplest models that
supports cluster formation is the modified Holstein model of eq.(\ref
{hamiltonian}). In the weak coupling regime the system is expected to form a
Luttinger liquid, while, for $\varepsilon \neq 0$ the effective electron
interaction, that is attractive for nearest neighbors, is responsible for
the formation of electron clusters at strong coupling. As we go from strong
to weak coupling, the cluster tends to break up and, in the intermediate
coupling regime and for phonon and electron energy scales not well separated
($\omega \simeq t$), all the electron configurations contribute to the
ground state without a dominant configuration (Fig.4). In this regime, that
is the most interesting both from the experimental and the theoretical point
of view, our calculation shows that different electron configurations can be
associated to very different lattice deformations. This result stems out
clearly from Fig.4 where we have plotted the LDPDF of two electron
configurations for $\lambda =1.2$ and $\varepsilon =0.1$.
\begin{figure}
\begin{center}
\epsfig{figure=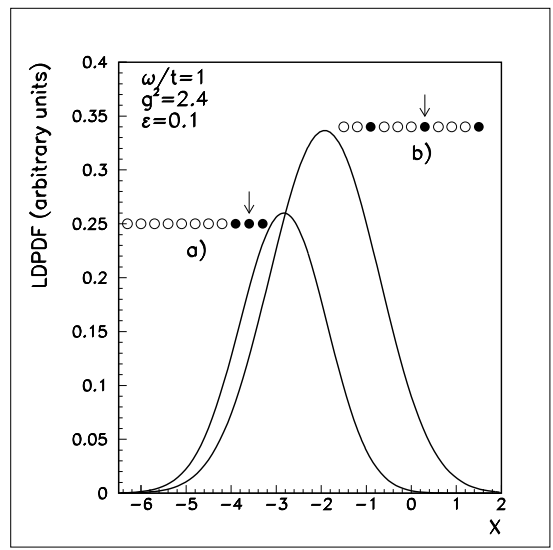,width=0.4\textwidth}
\end{center}
\caption{The LDPDF of the site indicated by the arrow for
two typical electron configurations are plotted
as a function of the lattice distortion in the intermediate coupling regime
for a system with nearest-neighbor e-ph interaction.
The lattice distortions are measured
in units of $(2M\omega)^{-1/2}$. The
system is made of three electrons on $L=11$ lattice sites.}
\label{Fig.4}
\end{figure}
 The site occupied
by the central electron in the cluster configuration (a) is characterized by
a LDPDF peaked at high values of deformation while the site occupied by the
central electron in configuration (b) presents much weaker deformations
typical of the weak coupling regime with larger spreading. The balance
between effective hopping, reduced by the cluster formation, and e-ph energy
gain can allow a larger lattice deformation energy.
The presence
of electron configurations with very different lattice deformations that
contribute to the ground state almost at same level suggests the existence
of significant quantum dynamical fluctuations triggered by e-ph
interactions. We think that our result, although limited to a simple 1D e-ph
model, provides support to a number of experimental evidences in cuprates
\cite{calvani,bilinge} and manganites\cite{edwards} where lattice distortions
have been observed by infrared, neutron scattering and X-ray spectroscopy.

{\em Conclusions. }In this report we have proposed a variational
approach, based on CS expansion, that is able to identify
optimized phonon basis useful for studying e-ph models.
The reliability of the model has been demonstrated for the extended
Holstein model emphasizing the role of quantum fluctuations.
In particular, it has been pointed out that very different lattice
distortions coexist at intermediate e-ph coupling.


\begin{thebibliography}{99}
\bibitem{nota1}  For one-dimensional systems, in the limit of infinite local
repulsion, the charge sector of the Hubbard model can be mapped onto a
spinless model.
\end{thebibliography}
\end{document}